# Coexistence of bulk-nodal and surface-nodeless Cooper pairings in a superconducting Dirac semimetal

Xian P. Yang[1*], Yigui Zhong[2], Sougata Mardanya[3], Tyler A. Cochran[1], Ramakanta Chapai[4], Akifumi Mine[2], Junyi Zhang[5*], Jaime Sánchez-Barriga[6, 7], Zi-Jia Cheng[1], Oliver J. Clark[6], Jia-Xin Yin[8], Joanna Blawat[4, 9], Guangming Cheng[10], Ilya Belopolski[1], Tsubaki Nagashima[2], Najafzadeh Sahand[2], Shiyuan Gao[5], Nan Yao[10], Arun Bansil[11], Rongying Jin[4, 9], Tay-Rong Chang[3, 12, 13], Shik Shin[2, 14, 15], Kozo Okazaki[2, 15, 16], M. Zahid Hasan[1, 10, 17†]

[1]Laboratory for Topological Quantum Matter and Advanced Spectroscopy (B7), Department of Physics, Princeton University, Princeton, NJ 08544, USA.
[2]Institute for Solid State Physics, University of Tokyo, Kashiwa, Chiba 277-8581, Japan.
[3]Department of Physics, National Cheng Kung University, Tainan 701, Taiwan.
[4]Department of Physics and Astronomy, Louisiana State University, Baton Rouge, LA 70803, USA.
[5]Institute for Quantum Matter and Department of Physics and Astronomy, Johns Hopkins University, Baltimore, MD 21218, USA.
[6]Helmholtz-Zentrum Berlin für Materialien und Energie, Elektronenspeicherring BESSY II, Albert-Einstein Strasse 15, Berlin 12489, Germany.
[7]IMDEA Nanoscience, C/ Faraday 9, Campus de Cantoblanco, Madrid 28049, Spain.
[8]Department of Physics, Southern University of Science and Technology, Shenzhen, Guangdong 518055, China.
[9]Center for Experimental Nanoscale Physics, Department of Physics and Astronomy, University of South Carolina, Columbia, SC 29208, USA.
[10]Princeton Institute for Science and Technology of Materials, Princeton University, Princeton, NJ 08544, USA.
[11]Department of Physics, Northeastern University, Boston, MA 02115, USA.
[12]Center for Quantum Frontiers of Research and Technology (QFort), Tainan 701, Taiwan.
[13]Physics Division, National Center for Theoretical Sciences, Taipei 10617, Taiwan.
[14]Office of University Professor, University of Tokyo, Kashiwa, Chiba 277-8581, Japan.
[15]Material Innovation Research Center, University of Tokyo, Kashiwa, Chiba 277-8581, Japan.
[16]Trans-scale Quantum Science Institute, University of Tokyo, Bunkyo-ku, Tokyo 113-0033, Japan.
[17]Lawrence Berkeley National Laboratory, Berkeley, CA 94720, USA.

*xiany@princeton.edu
*jzhan312@jhu.edu
†mzhasan@princeton.edu

**The interplay of nontrivial topology and superconductivity in condensed matter physics gives rise to exotic phenomena. However, materials are extremely rare where it is possible to explore the full details of the superconducting pairing. Here, we investigate the momentum dependence of the superconducting gap distribution in a novel Dirac material PdTe. Using high resolution, low temperature photoemission spectroscopy, we establish it as a spin-orbit coupled Dirac semimetal with the topological Fermi arc crossing the Fermi level on the (010) surface. This spin-textured surface state exhibits a fully gapped superconducting Cooper pairing structure below $T_c \sim 4.5$ K. Moreover, we find a node in the bulk near the Brillouin zone boundary, away from the topological Fermi arc. These observations not only demonstrate the band resolved electronic correlation between topological Fermi arc states and the way it induces Cooper pairing in PdTe, but also provide a rare case where surface and bulk states host a coexistence of nodeless and nodal gap structures enforced by spin-orbit coupling.**

Superconducting topological materials are emerging as the new frontier of condensed matter physics. Despite increasing interest in this field, few candidates have been identified because it is very challenging to confirm topological superconductivity experimentally [1-22]. While the topological surface states of three-dimensional (3D) topological insulators (TIs) can host topological superconductivity below the critical temperature ($T_c$), spin-momentum locked Fermi arcs in 3D Dirac semimetals could also induce topological superconductivity. Such a superconducting pairing between Fermi arcs with opposite spins and momenta has been proposed to reveal the nonlocal electronic correlations in topological semimetals [16]. However, the superconducting state of Fermi arcs has rarely been explored experimentally. Therefore, understanding the Cooper pairings of the Fermi arc states offers a new perspective for topological superconductors (TSCs) [23-27]. Moreover, recent theoretical work predicts the existence of nodes from the bulk bands in superconducting semimetals [28-30]. Thus, in superconducting semimetals that are also topological, surface Fermi arcs and nodal gap distribution of the bulk states can coexist, which in turn provide insight into the mechanism of unconventional and topological superconductivity.

Angle-resolved photoemission spectroscopy (ARPES) is arguably the most powerful technique for probing the superconducting gap distribution in the momentum space. Here, we use ARPES to demonstrate spin-orbit coupling (SOC) enforced surface-nodeless plus bulk-nodal pairings in a topological material PdTe. We first establish it as a 3D Dirac semimetal with a bulk Dirac point just below the Fermi level in its normal state with Fermi arc surface states on the (010) surface. Furthermore, our spin-resolved photoemission measurements indicate a polarized spin texture of these surface states in PdTe. Using high resolution ARPES, we observe a fully gapped Fermi arc below $T_c$, thus could potentially give rise to topological superconductivity. Moreover, we discover a node in the bulk bands near the Brillouin zone boundary below $T_c$. Therefore, PdTe displays a nodal gap in the bulk but nodeless gap distribution from the nontrivial surface states. Our study demonstrates a rare and exotic unconventional superconductor to study the interplay between nontrivial topology and superconductivity.

PdTe crystalizes in the NiAs type hexagonal structure with space group $P6_3/mmc$ [Fig. 1(a)]. Fig. 1(b) displays the Brillouin zone (BZ) where high symmetry points are indicated. For ARPES study, we focus on the (010) surface [$k_x$-$k_z$ surface in Fig. 1(b)]. Millimeter-sized PdTe single crystals were successfully synthesized. Powder X-ray diffraction (XRD) patterns [Fig. S1(a)] confirm that the sample has only a single phase with the hexagonal P6$_3$/mmc (194) structure [31]. The XRD pattern from the largest surface of a single crystal shows only (00n) peaks without any impurity peak, indicating the (001) direction [Fig. S1(b)]. The high quality of our single crystals is further confirmed by core level photoemission measurements with clear Pd d core levels [Fig. S1(c)]. Moreover, we identify a sharp superconducting transition in the electrical resistivity of PdTe with the critical temperature T$_c$ ~ 4.5 K [Fig. 1(c)], in agreement with previous reports [32, 33].

Like the well-established Dirac semimetal Na$_3$Bi [34], there is a type I bulk band crossing near the Fermi level along the Γ-A direction [Fig. 1(d)] in PdTe. Given the threefold rotational symmetry around the *c* axis [Fig. 1(b)], this band crossing remains gapless when spin-orbit coupling is included [34]. Since the two bulk Dirac crossings along the Γ-A and A-Γ will project onto the same position on the (001) surface in momentum space, it is impossible to experimentally observe the Fermi arc states on this surface. However, on the (010) surface, the two Dirac crossings are well separated in momentum, and the double Fermi arc surface states connecting the bulk Dirac points can be observed. Therefore, here, we focus on the (010) surface [Fig. 1(e)]. Indeed, the bulk band constant energy contour at the energy of the Dirac point [Fig. S2] clearly demonstrates the location of the Dirac crossings on (010). To help understand the connection of the surface Fermi arc states with the Dirac point, we calculate the surface spectrum of a semi-infinite Green's function projected on the (010), which displays the bulk Dirac cone as well as the Fermi arc surface states connecting to the Dirac crossing [Fig. 1(f)]. The connection of the surface Fermi arcs to the bulk Dirac points confirms PdTe as a Dirac semimetal.

Having identified the expected topology of PdTe, we systematically study its electronic structure and spin polarization at the (010) surface in its normal state. We demonstrate the topological Dirac fermion along the Γ-A direction and the corresponding Fermi arc states associated with the Dirac point. Fig. 2(a) shows the experimentally measured constant energy contour of PdTe at 100 meV binding energy corresponding to the binding energy of the Dirac point. The constant energy contour displays two bulk Dirac points and two Fermi arcs connecting the two Dirac points in the Brillouin zone (BZ). This is consistent with DFT calculations [Fig. 2(b)]. Overall, our ARPES data agree well with theoretical calculations [Fig. S3]. To better demonstrate the topology, we show the energy dispersion along the $k_x$ direction in cut 1 [Fig. 2(c)]. Several dispersive bands cross the Fermi level in Fig. 2(c)i. A comparison between (ii) and (iii) suggests that the inner dispersive band is a surface state. As shown in Fig. 2(a), cut 1 should pass through the Fermi arc states. Therefore, we conclude that the inner band is the Fermi arc surface state. Photon energy dependent ARPES data also verify the surface nature of this Fermi arc state [Fig. S4]. For further confirmation, we focus on cut 2 [Fig. 2(d)]. Since cut 2 passes through the bulk Dirac point as well as Fermi arc states, it can display the bulk-boundary correspondence. Indeed, (ii) and (iii) indicate the Dirac crossing below the Fermi level is a bulk state. Moreover, the bands emerging from the Dirac states have a surface origin, as they only appear in surface calculations. Our ARPES data in Fig. 2(d)i clearly show the surface states from the Dirac point. To observe the bulk Dirac crossing, we choose another direction, cut 3. Here, we can see the type I Dirac point through our data [Fig. 2(e)i], which agrees well with DFT calculations [Figs. 2(c)ii-iii]. By comparing ARPES and DFT in three cuts, we confirm the bulk Dirac nodes and the connecting Fermi arc surface states in PdTe. Apart from the Fermi arc surface states, there are also many topologically nontrivial surface states near the Fermi level [Fig. S5].

Next, we perform spin-resolved photoemission measurements to probe the spin polarization of the Fermi arc surface states. We measure two energy distribution curves (EDCs) indicated in Fig. 3(a). Since the two EDCs cut through the Fermi arc at the opposite side of Γ, they are expected to show reversed in-plane spin polarizations (along the $k_z$ direction). The spin-resolved EDCs and spin polarization curves [Figs. 3(b-c)] indeed display reversed spin polarizations, which confirm that the Fermi arc surface states are spin polarized in PdTe. The Fermi arcs might mix with bulk states, thus reducing the value of the experimentally measured spin polarization. As a comparison, the other two spin components along the $k_x$ and $k_y$ directions show negligible spin polarizations [Fig. S6].

To precisely determine the fermiology of the Fermi arcs across the transition temperature, we utilize low-temperature, high-resolution laser ARPES [35]. Our experimental data clearly reveal the curved Fermi arc surface state around the center of the BZ, in excellent agreement with theoretical calculations, as shown in Fig. 4(a). The shape of the Fermi arc states is slightly different from vacuum ultraviolet (VUV) ARPES results in Fig. 2 since laser ARPES measures a distinct surface termination [Fig. S7], but the nontrivial topology remains the same for both terminations [Fig. S8]. Having established the fermiology with laser ARPES, we show the superconducting gap in the momentum space for both the Fermi arc surface states and the bulk bands in PdTe. The positions of the correspondingly extracted EDCs are defined by an angle Ø in Fig. 4(a). For all the angles, the EDCs with the base temperature at around 2K show a small yet finite leading-edge shift compared to EDCs above the transition temperature [Fig. S9]. This finite edge shift across $T_c$ suggests a vanishing density of states at the Fermi level and thus displays the signature of a superconducting gap opening for both bulk and surface bands. For a better visualization, we symmetrize these EDCs with respect to the Fermi level in Figs. 4(b-c), and they all display a clear gap opening. Therefore, our low-temperature data visualize the formation of Cooper pairs inside a superconducting Fermi arc state in PdTe. This fully gapped Fermi arc may possibly lead to a topological superconducting phase in this 3D Dirac semimetal like the topological surface states in some iron-based superconductors do [5, 7]. Further theoretical works should address whether topological edge states can form in the potentially topological superconducting state in PdTe.

More intriguingly, despite the fully gapped bulk bands near the Fermi arc [Fig. 4(c)], we find that the bulk states near $\bar{Z}$ have a node in the superconducting state by exploring the bands close to the Brillouin zone boundary, away from the Fermi arc [Fig. 4(d)]. As shown in Fig. 4(e), the corresponding EDC displays no leading-edge shift across the critical temperature. Additionally, the symmetrized EDCs don't show any indication of a valley structure [Fig. 4(f)]. In other words, in contrast with the clear superconducting gap structures in Figs. 4(b-c), there is no sign of conventional superconductivity such as coherence peaks or a dip at the Fermi level in Fig. 4(f). Therefore, our measurements have uncovered a gapless node in the bulk bands, below the critical temperature.

Our results highlight a novel platform to study the interplay between superconductivity and topology. The spin-resolved ARPES data clearly demonstrate the spin-momentum locking of the Fermi arcs, which are fully gapped below $T_c$ as revealed by laser ARPES measurements.

Furthermore, the bulk bands have a node below $T_c$. So far, there are just a few photoemission studies on superconducting Dirac semimetal and none of them probes the superconducting Fermi arc states below $T_c$ [18-22]. Challenges are due to either the Dirac crossing far below the Fermi level [19, 21] or difficulty in measuring the side surface such as the (010) surface to directly observe the Fermi arc states by ARPES. Therefore, it is essential to search for an ideal 3D Dirac material system that (i) contains Dirac points near the Fermi level, (ii) has an accessible $T_c$ and (iii) can be cleaved for measuring superconducting Fermi arc states. PdTe is such a candidate satisfying all the above requirements. Moreover, it provides a rare case to study not only superconducting surface states already hard to access in other materials, but also is an uncommon demonstration of a gapless bulk node in a superconducting Dirac semimetal. The existence of a bulk node suggests the possibility of unconventional superconductivity in PdTe, thus distinguishing it from related superconducting semimetals like $PdTe_2$ [19, 20]. Our calculations suggest that this node is from the spin-orbital texture of the bulk Dirac crossing indicated in Fig. S10. Even with SOC, this bulk Dirac cone is gapless and protects the node near $\bar{Z}$. Our tight-binding calculations further confirm that the location of the nodal point should be close to the projection of this bulk Dirac cone in the momentum space [Fig. S11], consistent with experimental data. Thus, we demonstrate a SOC enforced bulk-nodal gap structure in PdTe along with a fully gapped Fermi arc surface state. Recent works [29, 30] suggest that for nodal bulk pairing in Dirac semimetal with $C_4$ rotational symmetry, high order or Majorana hinge modes can exist. It will be very interesting to observe these high order hinge modes experimentally. Similar states could also appear in PdTe with $C_3$ rotational symmetry, even though details may be different. Further theoretical and experimental studies are needed to verify the existence of potential high order hinge modes, thus leading to a better understanding of the complex intertwining of topology and superconductivity in PdTe.

**Acknowledgement**

The authors thank D. Lu and M. Hashimoto at Beamline 5-2 of the Stanford Synchrotron Radiation Lightsource (SSRL) at the SLAC National Accelerator Laboratory, CA, USA for support. The authors acknowledge enlightening discussions with X. Wu. Work at Princeton University is supported by the Gordon and Betty Moore Foundation (Grants No. GBMF4547 and No. GBMF9461; M. Z. H.). The ARPES work is supported by the United States Department of Energy (US DOE) under the Basic Energy Sciences program (Grant No. DOE/BES DE-FG-02-05ER46200; M. Z. H.). Materials characterization and the study of topological quantum properties are supported by the U.S. Department of Energy, Office of Science, National Quantum Information Science Research Centers, Quantum Science Center and Princeton University. Use of the Stanford Synchrotron Radiation Lightsource, SLAC National Accelerator Laboratory, is supported by the U.S. Department of Energy, Office of Science, Office of Basic Energy Sciences under Contract No. DE-AC02-76SF00515. Laser ARPES measurements in University of Tokyo are supported by Grants-in-Aid for Scientific Research (KAKENHI) (Grants No. JP19H01818 and No. JP19H00651) from the Japan Society for the Promotion of Science (JSPS) and by JSPS KAKENHI on Innovative Areas "Quantum Liquid Crystals" (Grants No. JP19H05826). We thank HZB for the allocation of synchrotron radiation beamtime at the U125-2-PGM beamline of BESSY II. J. S.-B. acknowledges financial support from the Impuls-und Vernetzungsfonds der Helmholtz-Gemeinschaft under grant No. HRSF-0067. Crystal growth and characterization (R. C., J. B., R. J.) are supported by NSF DMR-1504226. STEM characterization is performed with the use of Princeton University's Imaging and Analysis Center, which is partially supported by the Princeton Center for Complex Materials (PCCM), a National Science Foundation (NSF)-MRSEC program (DMR-2011750). T.-R. C. is supported by the Young Scholar Fellowship Program from the Ministry of Science and Technology (MOST) in Taiwan, under a MOST grant for the Columbus Program MOST110-2636-M-006-016, National Cheng Kung University, Taiwan, and National Center for Theoretical Sciences, Taiwan. This work is supported partially by the MOST, Taiwan, grant MOST107-2627-E-006-001. This research is supported in part by Higher Education Sprout Project, Ministry of Education to the Headquarters of University Advancement at National Cheng Kung University (NCKU). J. Z. and S. G. are supported as part of the Institute for Quantum Matter, an Energy Frontier Research Center funded by the U.S. Department of Energy, Office of Science, Basic Energy Sciences under Award No. DE-SC0019331. The work at Northeastern University is supported by the Air Force Office of Scientific Research under award number FA9550-20-1-0322, and it benefits from the computational resources of Northeastern University's Advanced Scientific Computation Center (ASCC) and the Discovery Cluster. T. A. C. acknowledges the support of the National Science Foundation Graduate Research Fellowship Program (DGE-1656466). I. B. acknowledges the generous support of the Special Postdoctoral Researchers Program, RIKEN during the late stages of this work. M. Z. H. acknowledges support from Lawrence Berkeley National Laboratory and the Miller Institute of Basic Research in Science at the University of California, Berkeley in the form of a Visiting Miller Professorship. M. Z. H. also acknowledges support from the U.S. Department of Energy, Office of Science, National Quantum Information Science Research Centers, Quantum Science Center.

X. P. Y., Y. Z., S. M., T. A. C., and R. C. contributed equally to this work.

**Figures**

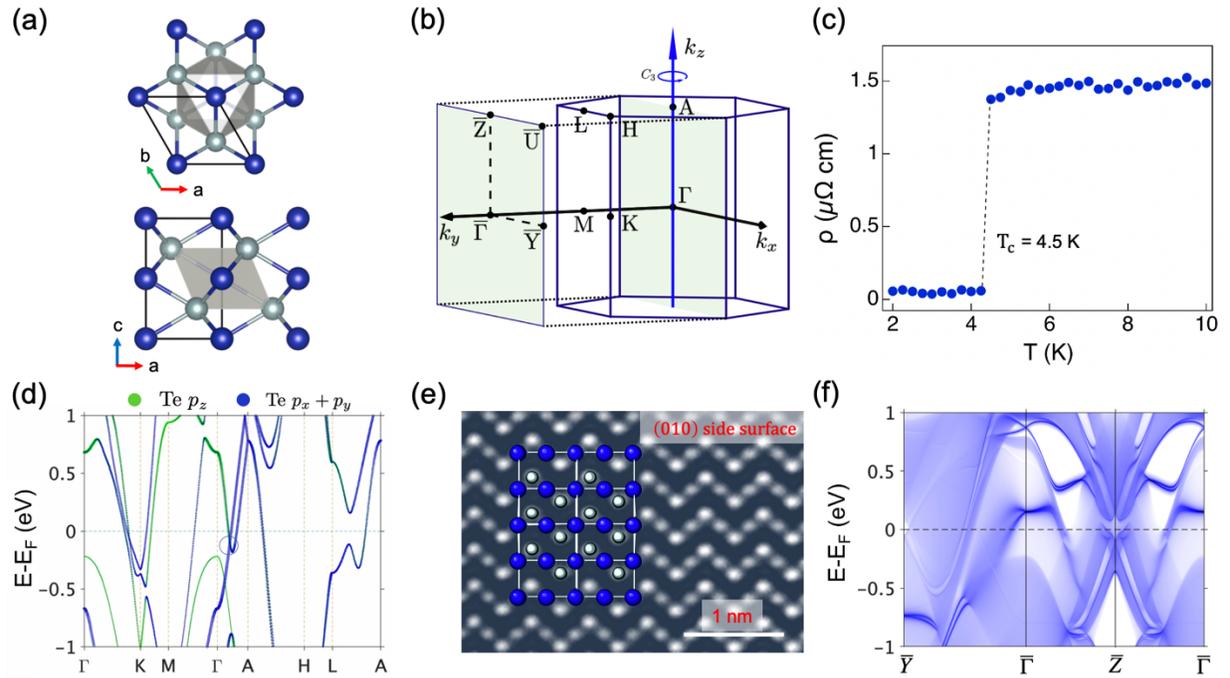

FIG. 1. Superconductivity and crystal structure of PdTe. (a) The crystal structure of PdTe. Blue is Pd atom and cyan is Te atom. (b) Bulk and surface Brillouin zones of PdTe. High symmetry points are marked. (c) Temperature-dependence zero-field electrical resistivity of PdTe at low temperature between 2 and 10 K. The dashed line is a guide for the eye. (d) Calculated bulk electronic structure of PdTe with spin-orbit coupling included. The bulk Dirac crossing is marked by the circle. The Fermi level is adjusted according to the experimental data. (e) Side surface scanning transmission electron microscope (STEM) image of PdTe. The unit cell is superimposed on top of the STEM image. (f) Semi-infinite Green's function surface calculations along the high symmetry directions on the (010) Pd-terminated surface.

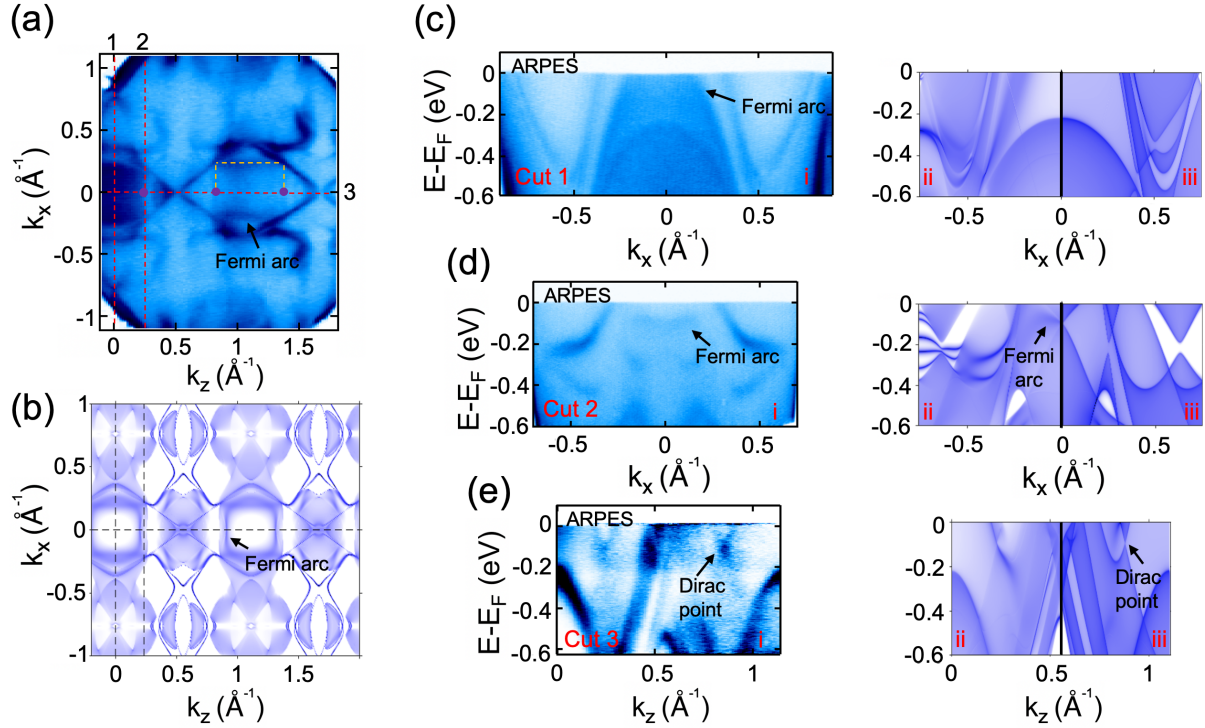

FIG. 2. Dirac crossing and Fermi arc state in PdTe above $T_c$. (a) ARPES constant energy contour at binding energy of 100 meV corresponding to the Dirac crossing on the Te-terminated (010) surface. Red dotted lines indicate ARPES dispersion cuts 1-3 in (c-e). Purple dots represent the projection of the Dirac cone and the yellow dotted lines connecting these dots are Fermi arc surface states. (b) Calculated constant energy contour of (a). (c) ARPES dispersion map (i) along cut 1 in (a). Black arrow is the Fermi arc state. Corresponding semi-infinite Green's function surface calculations including (ii) and excluding (iii) surface states contributions. (d) ARPES dispersion map (i) and corresponding DFT calculations (ii, iii) along cut 2 in (a). Black arrow in (i) indicates the surface state connecting to the bulk Dirac crossing. (e) ARPES dispersion map (i) and corresponding DFT calculations (ii, iii) along cut 3 in (a). Black arrow in (i) represents the bulk Dirac crossing.

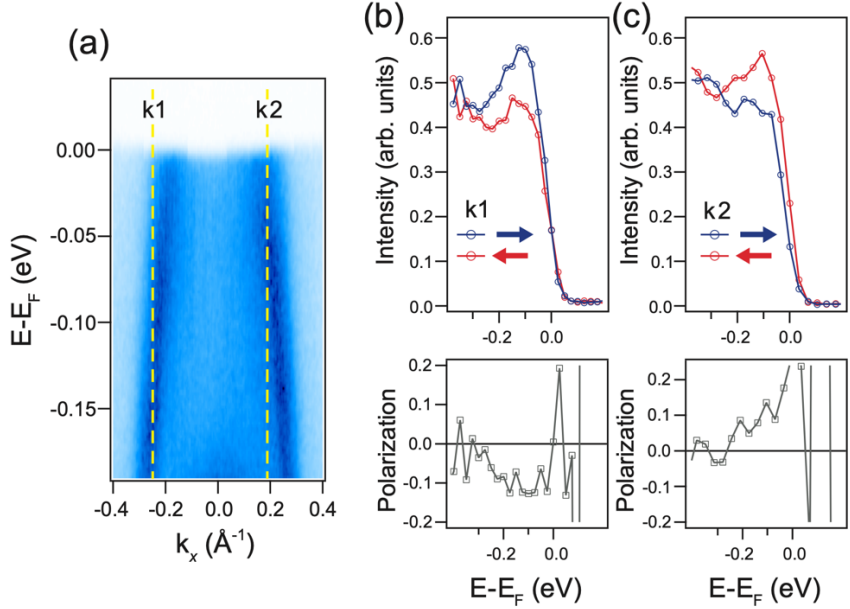

FIG. 3. Spin polarization of the Fermi arc state above $T_c$. (a) ARPES dispersion map of the Fermi arc state (same position as cut 1 in Fig. 2(a)). The yellow dotted lines represent the two momenta selected for spin-resolved photoemission measurements. (b-c) Spin-resolved intensity and polarization along the in-plane direction ($k_z$ direction) for k1 and k2.

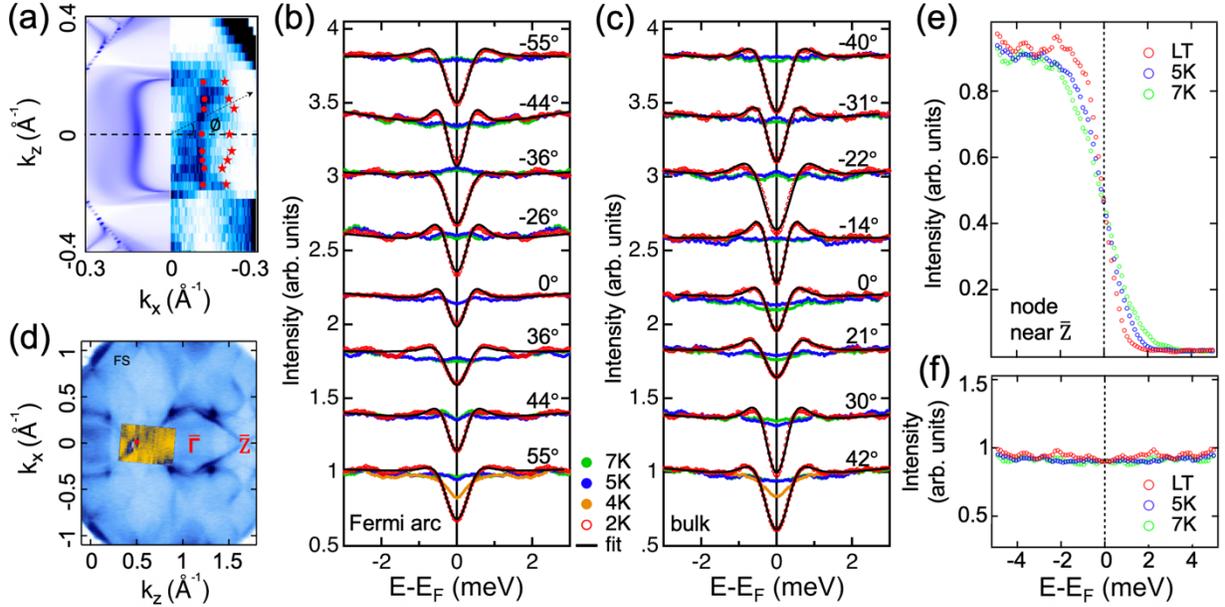

FIG. 4. Fully gapped surface states and gapless bulk node in the superconducting state of PdTe. (a) Comparison between laser ARPES Fermi surface map (right) and slab calculation (left) for the Pd termination in PdTe. Several points marked by red dots (red stars) along the Fermi arc (on the bulk states) are selected for superconducting gap measurements. ∅ defines the angles used in (b-c). (b-c) Symmetrized energy distribution curves (EDCs) corresponding to the red dots and stars in (a) below and above the critical temperature. The specific momentum locations of EDCs are marked by the Fermi surface angles defined in (a). All symmetrized EDCs reveal the opening of the superconducting gap below the critical temperature. Solid black lines represent fits to the Dynes' function. Experimental Fermi level position is determined by measuring a gold reference. (d) VUV ARPES Fermi surface map on the (010) surface. Laser ARPES Fermi surface map with much smaller momentum range is embedded on the VUV ARPES result. High symmetry points are marked in red. The location of the gapless node is indicated by the red square. (e) Temperature dependence of the EDC corresponding to the red square marked in (d) showing a node in the bulk band structure below the critical temperature. LT stands for the base temperature (~ 2 K) of our laser ARPES setup. (f) Symmetrized EDCs in (e) indicating the existence of a nodal point.